\documentclass[runningheads]{llncs}
\usepackage{graphicx}
\usepackage{amsmath}
\usepackage{mathtools}
\usepackage[ruled,vlined]{algorithm2e}
\usepackage{array}
\usepackage{cite}
\newcolumntype{C}[1]{>{\centering\let\newline\\\arraybackslash\hspace{0pt}}m{#1}}
\usepackage{multirow}
\usepackage{todonotes}
\usepackage{color}
\usepackage{ amssymb }
\usepackage{tablefootnote}
\usepackage{textcomp}

\usepackage{caption} 
\SetAlgorithmName{Alg.}{Alg.}{List of Algorithms} 
\SetAlCapNameFnt{\scriptsize}
\SetAlCapFnt{\scriptsize}

\begin{document}

\title{Fluid registration between lung CT and \\ stationary chest tomosynthesis images}
\titlerunning{CT to sDCT lung registration}

\author{Lin Tian\inst{1} \and Connor Puett\inst{2} \and Peirong Liu\inst{1} \and Zhengyang Shen\inst{1} 
\and Stephen R. Aylward\inst{3} \and Yueh Z. Lee\inst{2} \and Marc Niethammer\inst{1}}
\authorrunning{L. Tian et al.}

\institute{Department of Computer Science, University of North Carolina at Chapel Hill \and
Department of Radiology, University of North Carolina at Chapel Hill \and
Kitware, Inc.}

\maketitle

\begin{abstract}
Registration is widely used in image-guided therapy and image-guided surgery to estimate spatial correspondences between organs of interest between planning and treatment images. However, while high-quality computed tomography (CT) images are often available at planning time, limited angle acquisitions are frequently used during treatment because of radiation concerns or imaging time constraints. This requires algorithms to register CT images based on limited angle acquisitions. We, therefore, formulate a 3D/2D registration approach which infers a 3D deformation based on measured projections and digitally reconstructed radiographs of the CT. Most 3D/2D registration approaches use simple transformation models or require complex mathematical derivations to formulate the underlying optimization problem. Instead, our approach entirely relies on differentiable operations which can be combined with modern computational toolboxes supporting automatic differentiation. This then allows for rapid prototyping, integration with deep neural networks, and to support a variety of transformation models including fluid flow models. We demonstrate our approach for the registration between CT and stationary chest tomosynthesis (sDCT) images and show how it naturally leads to an iterative image reconstruction approach.
\end{abstract}

\section{Introduction}
Image registration is an enabling technology for image-guided interventional procedures (IGP)\cite{jaffray2007review}, as it can determine spatial correspondences between pre- and intra-intervention images. Conventionally, pre-interventional images are 3D CT or magnetic resonance (MR) images. Intra-interventional images are typically 2D projective X-rays or ultrasound images~\cite{markelj2012review}. While fast, cheap, and easy to use, these intra-interventional images lack anatomical features and spatial information. It is therefore desirable to combine high-quality 3D pre-intervention images with sDCT~\cite{shan2014stationary}. In sDCT, a set of projections is acquired from spatially distributed carbon nanotube-enabled X-ray sources within a limited scanning angle. Given the stationary design, the projections can be acquired fast with high resolution and provide more spatial information than individual X-ray projections. However, this immediately raises the question of how to register a 3D CT to an sDCT volume or its 2D projections.

This may be accomplished by two approaches: (1) reconstructing a 3D image from the 2D projections and proceeding with standard 3D to 3D registration; (2) computing image similarities in the projection space (i.e., by also computing projections from the 3D volume) from which a 3D transformation is inferred. While conceptually simple, the first approach may result in dramatically different 3D image appearances as 3D reconstructions from a small set of projections with a limited scanning angle range will appear blurry and will smear anatomical structures (e.g., the ribs across the lung). Similar appearance differences have been addressed in~\cite{tomazevic20053} for rigid transformations via an information-theoretic similarity measure. The second approach is conceptually more attractive in a limited angle setting, as it {\it simulates} the limited angle acquisitions from the high-quality 3D image. Hence, sets of projection images with {\it similar} appearance can be compared. Such an approach is called 3D/2D registration.  
Different 3D/2D registration approaches have been proposed~\cite{markelj2012review}, but most only address rigid or affine transformations~\cite{tomazevic20053, fu2008fast, aouadi2008accurate,jonic2003multiresolution,jans20063d}. Less work considers more flexible transformations or even non-parametric models for 3D/2D registration~\cite{flach2014deformable,prummer2006multi}. A particular challenge is the efficient computation of gradients with respect to the transformation parameters. Zikic et al.~\cite{zikic2008deformable} establish correspondences between vessels in 3D and a single 2D projection image using a diffusion registration model~\cite{modersitzki2004numerical} by exploiting vessel geometry. Pr\"ummer et al.~\cite{prummer20052d,prummer2006multi} use curvature registration~\cite{modersitzki2004numerical} and a similarity measure in projection space based on algebraic image reconstruction. 
The work by Flach et al.~\cite{flach2014deformable} relates most closely to our approach. It is based on Demons registration and digitally reconstructed radiographs (DRR) 
to compute image similarity in projection space. They propose a sophisticated optimization approach relying on explicit gradient computations. However, gradient computations quickly become unwieldy for complex transformation models which might explain the dearth of 3D/2D models for non-parametric registration. This complexity also hampers flexible modeling and rapid prototyping. Further, while some of the approaches above consider limited numbers of projections, they do not address limited angle acquisitions.  

Our starting point for 3D/2D registration is a DRR model for the stationary design of sDCT, which we use to compute image similarity in projection space. In contrast to previous approaches, all our components (the registration model, the similarity measure, and the DRR computation) support automatic-differentiation~\cite{griewank2008evaluating,NEURIPS2019_9015}. This allows focusing our formulation on the forward model and supports the flexible use of our DRR component. Specifically, within one framework we can support a variety of parametric and non-parametric registration models and can also formulate an iterative image reconstruction approach.

\paragraph{\bf{Contributions:}} 1) We propose a differentiable ray-cast DRR generator consistent with sDCT; 2) We propose a 3D/2D registration framework based on these DRR projections which support automatic differentiation. This allows us to easily integrate various registration models (including non-parametric ones) without the need for complex manual gradient calculations; 3) We show how our approach allows for a straightforward implementation of a limited angle image reconstruction algorithm; 4) We demonstrate the performance of our approach for CT to simulated sDCT registration and explore its performance when varying the number of projections and the range of scanning angles.

\section{Ray-cast DRR generator}
\label{sec:ray-cast}

\begin{figure}[htb]
	\centering
	\begin{minipage}[t]{0.45\linewidth}
		\centering
		\hspace{1cm}
		\includegraphics[height = 0.45\linewidth]{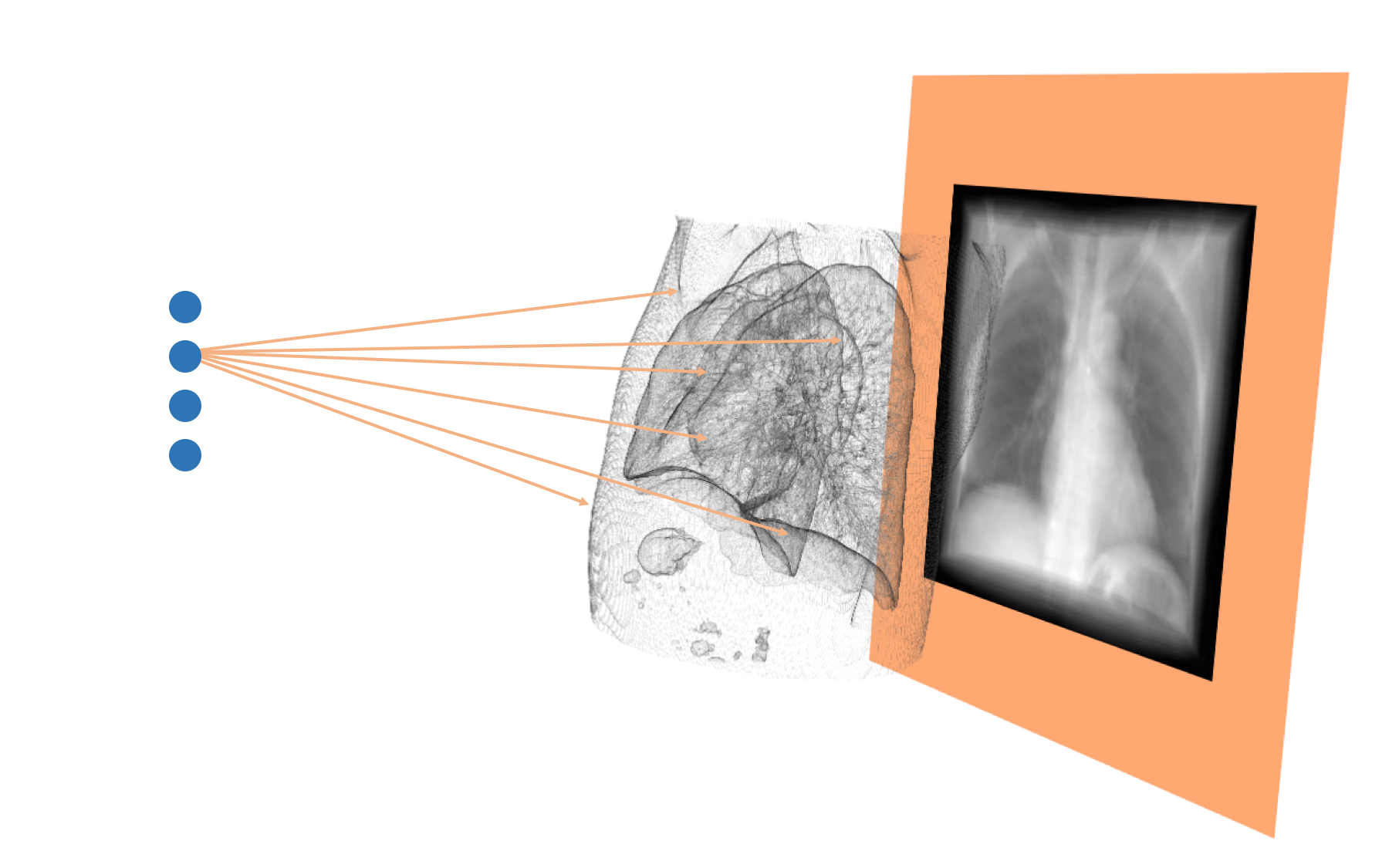}
	\end{minipage}
	\begin{minipage}[t]{0.45\linewidth}
		\centering
		\hspace{-1cm}
		\includegraphics[height = 0.45\linewidth]{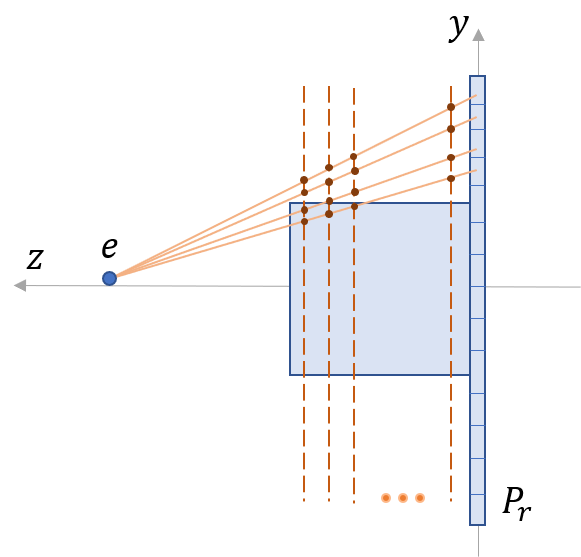}
	\end{minipage}
	\caption{Illustration of sDCT and our DRR generation. Left: Emitters are spatially distributed in sDCT~\cite{shan2014stationary}. Right: Side view
	of DRR geometry.
	}\label{fig:projection}
\end{figure}

Given a high-quality volumetric image $I_0(x),~x\in{R^3}$ (e.g., based on CT) and one or a set of projection images $I_1(y),~y\in{R^2}$, we generate DRRs by shooting virtual rays through the 3D volume to a virtual receiver plane $P_r$ of size $W\times{H}$ (Fig.~\ref{fig:projection}). 
Assume we are in the coordinate system of the receiver plane whose center is placed and will stay at the origin $(0,0,0)$ for all the emitters according to the stationary design of sDCT. The position of points in $P_r$ can then be represented as $p_r=\{(w,h,0)^T|\frac{-{W}}{2}\leq{w}\leq\frac{W}{2},\, \frac{-H}{2}\leq{h}\leq\frac{H}{2}\}$. Given an emitter at position $e\in{R^3}$ and a point $x$ on the ray from $e$ to $p_r$, we represent the ray and DRR of this geometry respectively as
\begin{equation}
    x = e + \lambda\hat{r},\quad \hat{\vec{r}}=\vec{r}/\|\vec{r}\|_2,\quad \vec{r}=p_r-e,\quad
    I_{proj} = \int{I_0(x)~dx}. \label{equ_ray_drr}
\end{equation}

We discretize the integral along the $z$ axis of the receiver plane (Fig.~\ref{fig:projection}) and obtain planes $P=z$ parallel to the receiver plane. Given the point $x_{z}=(0,0,z)^T$ on the plane with the plane normal $\vec{n}=(0,0,1)^T$, any point $x$ on the plane $P=z$ can be written as 
\begin{equation}
     (x-x_{z})\cdot{\vec{n}}=0. \label{equ_plane}
\end{equation}

Combining Eqs.~\ref{equ_ray_drr} and~\ref{equ_plane}, 
we obtain the intersection between the ray and the plane $P=z$:
\begin{equation}
    x = e+\frac{(x_{z}-e)\cdot\vec{n}}{\vec{r}\cdot\vec{n}}\hat{\vec{r}},\,
    \label{equ_transform_matrix}
\end{equation}
The DRR ($I_{proj}$) can then be written as
\begin{equation}
    I_{proj}(w,h,e) = \sum^Z_{z=0}{I_0(e+\frac{(x_{z}-e)\cdot\vec{n}}{\vec{r}\cdot\vec{n}}\hat{\vec{r}})~dz},\ \vec{r}=(w,h,0)^T-e,\ dz=||\frac{\vec{r}}{\vec{r}\cdot\vec{n}}||_2 .\ \label{equ_proj}
\end{equation}
It is possible to write Eq.~\ref{equ_proj} in tensor form where all rays (for all receiver plane coordinates $(w,h)$) are evaluated at once. This allows efficient computation on a GPU and only involves tensor products, interpolations, and sums. Hence, we can also efficiently implement derivatives of a DRR with respect to $I_0$ using automatic differentiation. This allows easy use of DRRs in general loss functions.

\section{Registration model}
\label{sec:lddmm}

Our goal is to develop a generic 3D/2D registration approach that can be combined with any transformation model. In general, image registration is formulated as an energy minimization problem of the form~\cite{modersitzki2004numerical}
\begin{equation}
    \theta^*=\arg\min_\theta\, {\rm Reg}(\theta) + \lambda\,{\rm Sim}( I_0\circ\Phi^{-1}_\theta,\, I_1),~\lambda>0,
    \label{eq:standard_registration}
\end{equation}
where the transformation $\Phi^{-1}_\theta$ is parameterized by $\theta$, $\lambda$ is a weighting factor, ${\rm Reg}(\theta)$ is a regularizer on the transformation parameters, ${\rm Sim}(A,B)$ is a similarity measure (e.g., sum of squared differences, normalized cross correlation, or mutual information) between two images $A$, $B$ of the same dimension, and $I_0$, $I_1$ are the source and target images respectively. Since we are interested in 3D/2D registration, we instead have a 3D source image $I_0$ and a set of projection target images $\{I_1^i\}$. The similarity measure compares the DRRs of the 3D source image to their corresponding projection images. This changes Eq.~\ref{eq:standard_registration} to
\begin{equation}
    \theta^* = \arg\min_\theta {\rm Reg}(\theta) + \frac{\lambda}{n}\sum_{i=1}^n {\rm Sim}(P^i[I_0\circ\Phi_\theta^{-1}],I_1^i),~\lambda>0,\label{eq:projection_registration}
\end{equation}
where we now average over the similarity measures with respect to all the $n$ projection images, and $P^i[\cdot]$ denotes the projection consistent with the geometry of the $i$-th projection image $I_1^i$ (i.e., its DRR). Note that neither the regularizer nor the transformation model changed compared to the registration formulation of Eq.~\ref{eq:standard_registration}. The main difference is that the 3D transformed source image $I_0\circ\Phi_\theta^{-1}$ undergoes different projections and the similarity is measured in the projection space. Clearly, optimizing over such a formulation might be difficult as one needs to determine the potentially rather indirect dependency between the transformation parameter $\theta$, and the image differences measured after projection. While such dependencies have previously been determined analytically~\cite{flach2014deformable,prummer2006multi}, they make using advanced registration models and rapid prototyping of registration approaches difficult. Instead, we base our entire registration approach on automatic differentiation. In such a framework, changing the transformation model becomes easy. We explore an affine transformation model ($\Phi^{-1}_\theta(x)=Ax+b$, where $\theta=\{A,b\}$,\, $A\in\mathbb{R}^{3\times3}$,\, $b\in\mathbb{R}^{3\times 1}$), which has previously been used in 3D/2D registration. To capture large, localized deformations we also use a fluid-registration model after affine pre-registration. We use the large deformation diffeomorphic metric mapping (LDDMM) model~\cite{miller2002metrics,beg2005computing}, where a spatial transformation is parameterized by a spatio-temporal velocity field. The LDDMM formulation of 3D/2D registration in its shooting form~\cite{singh2013vector,vialard2012diffeomorphic,shen2019region} becomes
\begin{eqnarray}
    &\theta^*(x)& = \arg\min_{\theta(x)} \langle \theta, K \star \theta\rangle + \frac{\lambda}{n}\sum_{i=1}^n {\rm Sim}(P^i[I_0\circ\Phi^{-1}(1)],I_1^i),~\lambda>0,\label{eq:lddmm_energy}\\
    &\Phi^{-1}_t& + D\Phi^{-1}v=0,~\Phi^{-1}(0,x)=\Phi^{-1}_0(x),\label{eq:phi_advect}\\
    &m_t& + div(v)m + Dv^T(m)+Dm(v)=0,~v=K\star m,~m(x,0) = \theta(x),\label{eq:epdiff}
\end{eqnarray}
where $\theta$ is the initial momentum vector field which parameterizes the transformation over time via an advection equation (Eq.~\ref{eq:phi_advect}) on the transformation map $\Phi^{-1}$ controlled by the spatially-varying velocity field, $v$, whose evolution is governed by the Euler-Poincar\'e equation for diffeomorphisms (EPDiff; Eq.~\ref{eq:epdiff}); $\Phi^{-1}(1)$ is the transformation at time $t=1$ which warps the source image and $\Phi^{-1}_0(x)$ is its initial condition (set to the result of our affine pre-registration); $K$ is a regularizer (we choose a multi-Gaussian~\cite{risser2010simultaneous}). This registration model can capture large deformations (as occur in lung registration), is parameterized entirely by its initial momentum field, and can guarantee diffeomorphic transformations. While technically possible, it is easy to appreciate that taking derivatives of the LDDMM 3D/2D registration model is not straightforward. However, our implementation based on automatic differentiation only requires implementation of the forward model (Eqs.~\ref{eq:lddmm_energy}-\ref{eq:epdiff}). Gradient computation is then automatic.

Alg.~\ref{algo_lddmm} describes our 3D/2D registration algorithm:

\begin{algorithm}[H]\label{algo_lddmm}
\scriptsize
\SetAlgoLined
\KwResult{$\theta^*,\Phi^{-1}_{afffine}$}
Initialize $v=\boldsymbol{0}$ and $\Phi^{-1}(0)=\Phi^{-1}_{affine}$\;
\While{loss not converged}{
Compute $\phi^{-1}$\ and warp $I_0$ using $\phi^{-1}$\;
Compute projection images $\{P^i[I_0\circ\Phi^{-1}]\}$ using tensor form of Eq.~\ref{equ_proj}\;
Compute ${\rm Sim}_{NGF}(P^i[I_0\circ\Phi^{-1}], I_1^i)$ and ${\rm Reg}(\theta)$\;
Back-propagate to compute $\nabla_\theta loss$ and update $\theta$
}
\caption{3D/2D registration}
\end{algorithm}

\section{Limited angle image reconstruction}
\label{sec:reconstruction}

We have shown how our DRR model integrates within our 3D/2D registration formulation. Now we 
apply it for an iterative reconstruction algorithm (Alg.~\ref{algo_reconstruction}). 

Given a set of $n$ 2D projection images $\{I^i\}$, our goal is to determine the 3D image $I$ which is most consistent with these projections. Specifically, we solve the following minimization problem:
\begin{equation}
    I^*=\arg\min_I{\frac{1}{n}\sum_{i=1}^n {\rm Sim}(P^i[I],I^i)+\lambda_1 {\rm ReLU}(-I)+\frac{\lambda_2}{|\Omega|}\int_\Omega ||\nabla{I}||_2~dx},~\lambda_1,\lambda_2\geq 0,\label{equ_reconstruction}
\end{equation}
where $P^i[\cdot]$ is the DRR generated for the i-th emitter (Eq.~\ref{equ_proj}). ${\rm Sim}$ is the $L_1$ loss between $P^i[I]$ and $I^i$. We add two regularization terms: 1) ${\rm ReLU}(I)$, which encourages positive reconstructed values, and 2) $\int\|\nabla I\|_2~dx$, which measures the total variation of $I$,  encouraging piece-wise constant reconstructions.  

\begin{algorithm}[H]\label{algo_reconstruction}
\scriptsize
\SetAlgoLined
\KwResult{$I^*$}
Initialize $I=\boldsymbol{0}$\;
\While{loss not converged}{
Compute $\{P^i[I]\}$ using tensor form of Eq.~\ref{equ_proj}\;
Compute ${\rm Sim}(P^i[I], I^i)$, $ReLU(-I)$ and $\int ||\nabla{I}||_2~dx$\;
Back-propagate to compute $\nabla_I loss$ to update $I$
}
\caption{Iterative reconstruction}
\end{algorithm}

\section{Experiments and results}
\label{sec:experiments}

\subsubsection{Dataset}

We use two lung datasets for evaluation: 1) a paired CT-sDCT dataset including raw sDCT projections, 2) a 3D CT dataset with landmarks.

\paragraph{CT-sDCT:}
We acquired 3 pairs of CT/sDCT images including 31 sDCT projections for each sDCT image with uniformly spaced emitters covering a scanning range of 12 degrees. We use this dataset to test our reconstruction approach.

\paragraph{DIR-Lab 4DCT:}
This dataset contains 10 pairs of inhale/exhale lung CT images with small deformations and 300 corresponding landmarks per pair. 
We use this dataset to synthesize corresponding 3D/2D image sets. Specifically, we synthesize projection images from the inspiration phase CT (via our DRR model) and treat them as the target images. As we have the true 3D images with landmarks we can then perform quantitative validation.

\subsection{sDCT reconstruction}
\label{sec:exp_sdct}
We compare our reconstruction result (using Alg.~\ref{algo_reconstruction} and the sDCT projections) to the given sDCT reconstructions (based on~\cite{wu2015adapted}) of the CT-sDCT dataset.

\begin{figure}[htb]
    \centering
    \includegraphics[width=\textwidth]{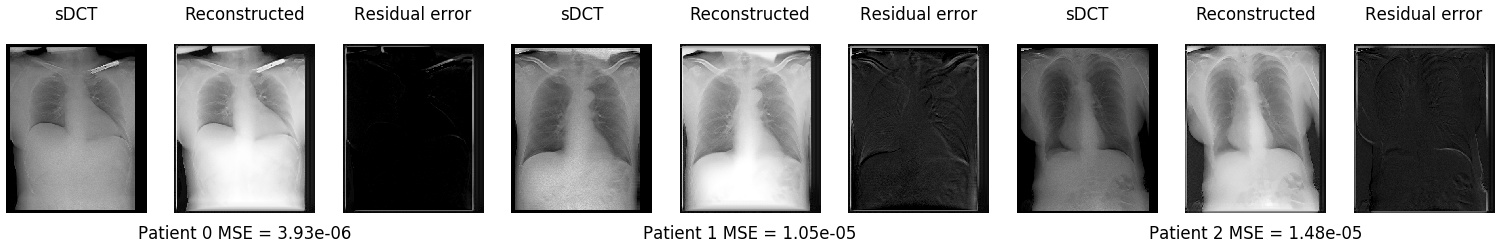}
    \caption{Comparisons between our reconstructed volumes (Alg.~\ref{algo_reconstruction}) and sDCT volumes (we only show one slice per case). Columns 1, 3, 6: provided sDCT images; columns 2, 5, 8: reconstructed images from our method; columns 3, 6, 9: residual error computed by subtracting our reconstructed image from the sDCT image.}
   
    \label{fig:reconstruction}
\end{figure}

Fig.~\ref{fig:reconstruction} shows the resulting reconstructions. The mean squared error (MSE) between our reconstruction algorithm and the reconstruction algorithm of~\cite{wu2015adapted} is small, demonstrating that our method recovers image structure equally well. This result also qualitatively demonstrates the correctness of our DRR projection algorithm.

\subsection{3D/2D registrations}\label{sec:exp_dirlab}
\label{subsec:synthetic-3d-2d-registration}

Given a CT image and a set of projection images we follow two 3D/2D registration approaches: 1) our approach (Alg.~\ref{algo_lddmm}) and 2) 3D/3D registration by first reconstructing a 3D volume (Alg.~\ref{algo_reconstruction}) from the set of projection images. The second approach is our baseline demonstrating that the vastly different appearances of the reconstructed 3D volumes (for a low number of projections) and the CT images result in poor registration performance, which can be overcome using our approach. To quantitatively compare both approaches, we use the DIR-Lab lung CT dataset to create synthetic CT/projection set pairs, where we keep the inspiration phase CT image and create the synthetic DRRs from the expiration phase CT (which constitute our simulated raw sDCT projections). While sDCT imaging is fast, the imaging quality is lower than for CT. 
For comparison, we provide 3D-CT/3D-CT registration results from the literature.

\begin{table}[htb]
\caption{Mean distance of expert landmarks before and after registration with our proposed 3D/2D methods and comparison to a SOTA 3D-CT/3D-CT method.}\label{tab_synthetic_exp}
\scriptsize
\begin{tabular}{l|C{1cm}|C{1.7cm}|C{1.7cm}|C{1.7cm}|C{1.7cm}|C{1.7cm}}
\hline

\multicolumn{1}{c|}{\multirow{3}{*}{Dataset}} & \multicolumn{1}{c|}{\multirow{3}{*}{Initial}} & \multicolumn{4}{c|}{3D-2D registration task} & \multicolumn{1}{c}{3D-3D} \\ \cline{3-6}
\multicolumn{1}{c|}{} & \multicolumn{1}{c|}{} & \multicolumn{2}{c|}{Reconstruction} & \multicolumn{2}{c|}{Projection} &\multicolumn{1}{c}{reg. 
task}  \\ \cline{3-7} 
\multicolumn{1}{c|}{} & \multicolumn{1}{c|}{} & Affine & LDDMM & Affine & LDDMM & \multicolumn{1}{c}{pTVreg\cite{vishnevskiy2017isotropic}} \\ \hline
4DCT1 &  4.01 & 10.14 & 19.51  & 2.99 & 1.74   & 0.80\\
4DCT2 &  4.65 & 19.67 & 29.71 & 2.10 & 1.74   & 0.77\\
4DCT3 &  6.73 & 6.07 & 14.90 & 3.11 & 2.24   & 0.92\\
4DCT4 &  9.42 & 24.02 & 29.24 & 3.37 & 2.54  & 1.30\\
4DCT5 &  7.10 & 19.79 & 26.21 & 4.48 & 2.88  & 1.13\\
4DCT6 &  11.10 & 21.81 & 30.51 & 6.13 & 5.18 & 0.78\\
4DCT7 &  11.59 & 16.83 & 26.63 & 4.64 & 4.07  & 0.79\\
4DCT8 &  15.16 & 14.78 & 18.85 & 7.85 & 2.73  & 1.00\\
4DCT9 &  7.82 & 24.44 & 29.04 & 3.70 & 3.76  & 0.91\\
4DCT10 &  7.63 & 16.50 & 26.28 & 3.57 & 2.63  & 0.82\\\hline
Mean & 8.52 & 17.40 & 25.08 & 4.19 & 2.95  & 0.92\\
Std. & 3.37 & 5.89 & 5.40 & 1.69 & 1.08  & 0.17\\
\hline
\end{tabular}
\centering
\end{table}

\paragraph{Registration of 3D-CT/synthetic 2D sDCT projections} We resample both expiration/inspiration lung CT images to $1\times{1}\times{1}$~mm. To make our results comparable to state-of-the-art (SOTA) 3D-CT/3D-CT lung registration approaches, we segment the lung and remove all image information outside the lung mask. 
Then, we create DRRs from the expiration phase lung CT to simulate a set of sDCT projection images with the same emitter geometry as in our CT-sDCT dataset~\cite{shan2014stationary}. However, we use only 4 emitters placed evenly within 11 degrees (compared to 31 emitters in CT-sDCT dataset) 
to assess registration behavior for very few 
projections. We use $\lambda = 0.5$, the normalized gradient field (NGF) similarity measure~\cite{haber2006intensity}, stochastic gradient descent optimization for affine registration with $\lambda = 0.01$; and NGF, lBFGS~\cite{liu1989limited} for LDDMM registration.  
Tab.~\ref{tab_synthetic_exp} shows the registration results. As expected, our affine and LDDMM projection results improve over the initial alignment and LDDMM improves over affine registration, indicating more complex transformations can be recovered with our 3D/2D registration approach. 
Due to the limited number of projections our results are not as accurate as direct 3D-CT/3D-CT registration (e.g., SOTA pTVreg~\cite{vishnevskiy2016isotropic}), which, however, would not be practical in an interventional setting. 

\paragraph{Registration of 3D-CT/synthetic sDCT reconstruction} We apply Alg.~\ref{algo_reconstruction} to reconstruct the sDCT from the synthetic projection images of the 3D-CT expiration phase image. 
We set $\lambda_1=100,\,\lambda_2=1$ in Eq.~\ref{equ_reconstruction}, then register the 3D-CT at inhale to the reconstructed sDCT using 3D/3D LDDMM registration.

Tab.~\ref{tab_synthetic_exp} shows that 3D/2D registration outperforms 3D/3D registration via the reconstructed sDCT. This is because of the vast appearance difference of the sDCT image (compared to 3D-CT), because of limited-angle reconstruction, and because of the low number of projections. In fact, when using the reconstructed sDCT, registration results are worse (for affine and LDDMM) than not registering at all. Hence, using the 3D/2D approach is critical in this setting.

\subsection{Influence of angle range and number of projections}

We explore how our 3D/2D registration results change with the number of projections (fixing angle range of 11 degrees) and with the angle range (fixing the number of projections at 4). Results are based on DIR-Lab case 4DCT5 (Tab.~\ref{tab_synthetic_exp}).

\begin{figure}[htb] 
\centering
  \begin{minipage}[b]{0.248\textwidth}
    \centering
    \includegraphics[width=1\textwidth]{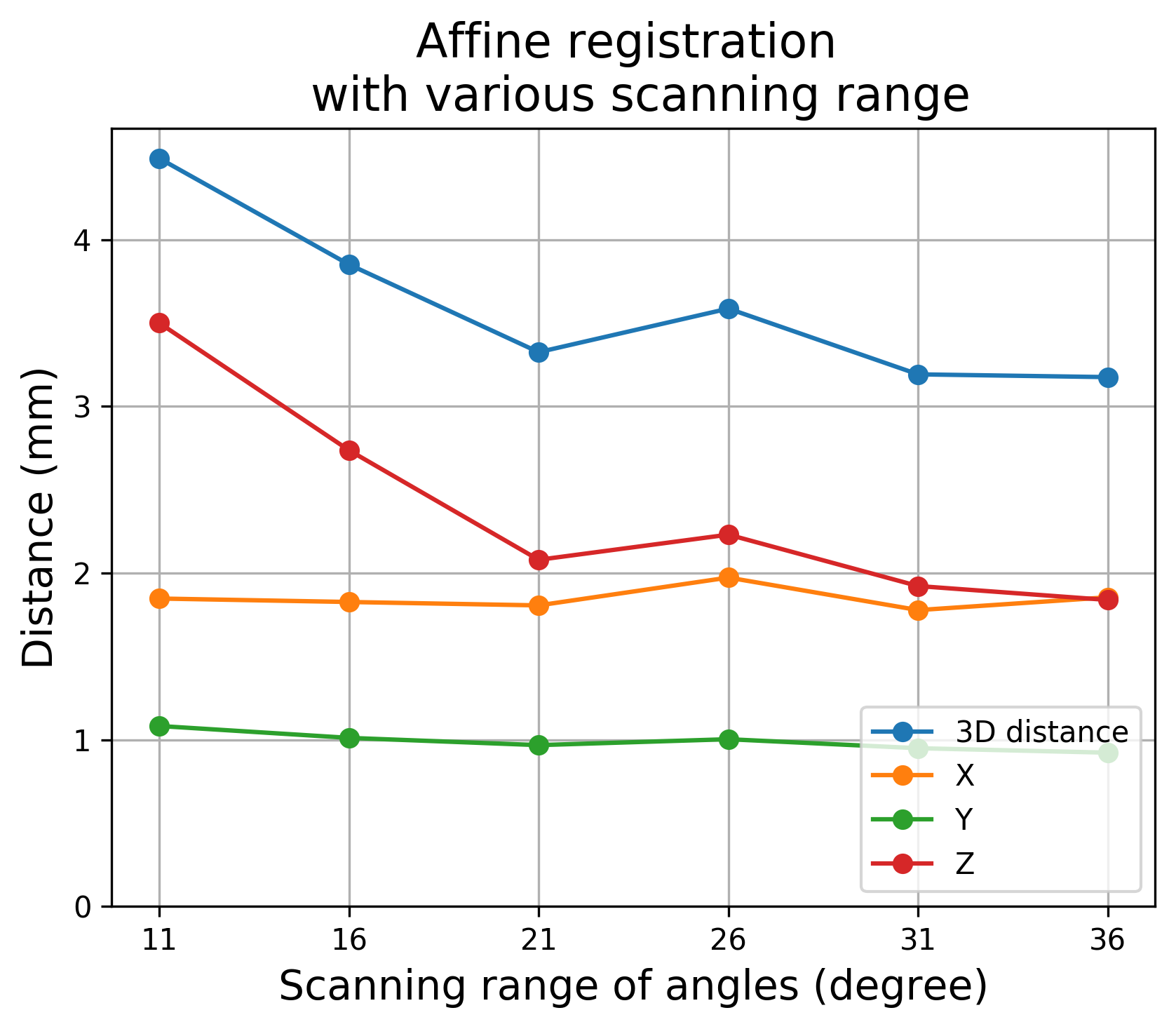} 
  \end{minipage}
  \begin{minipage}[b]{0.248\textwidth}
    \centering
    \includegraphics[width=1\textwidth]{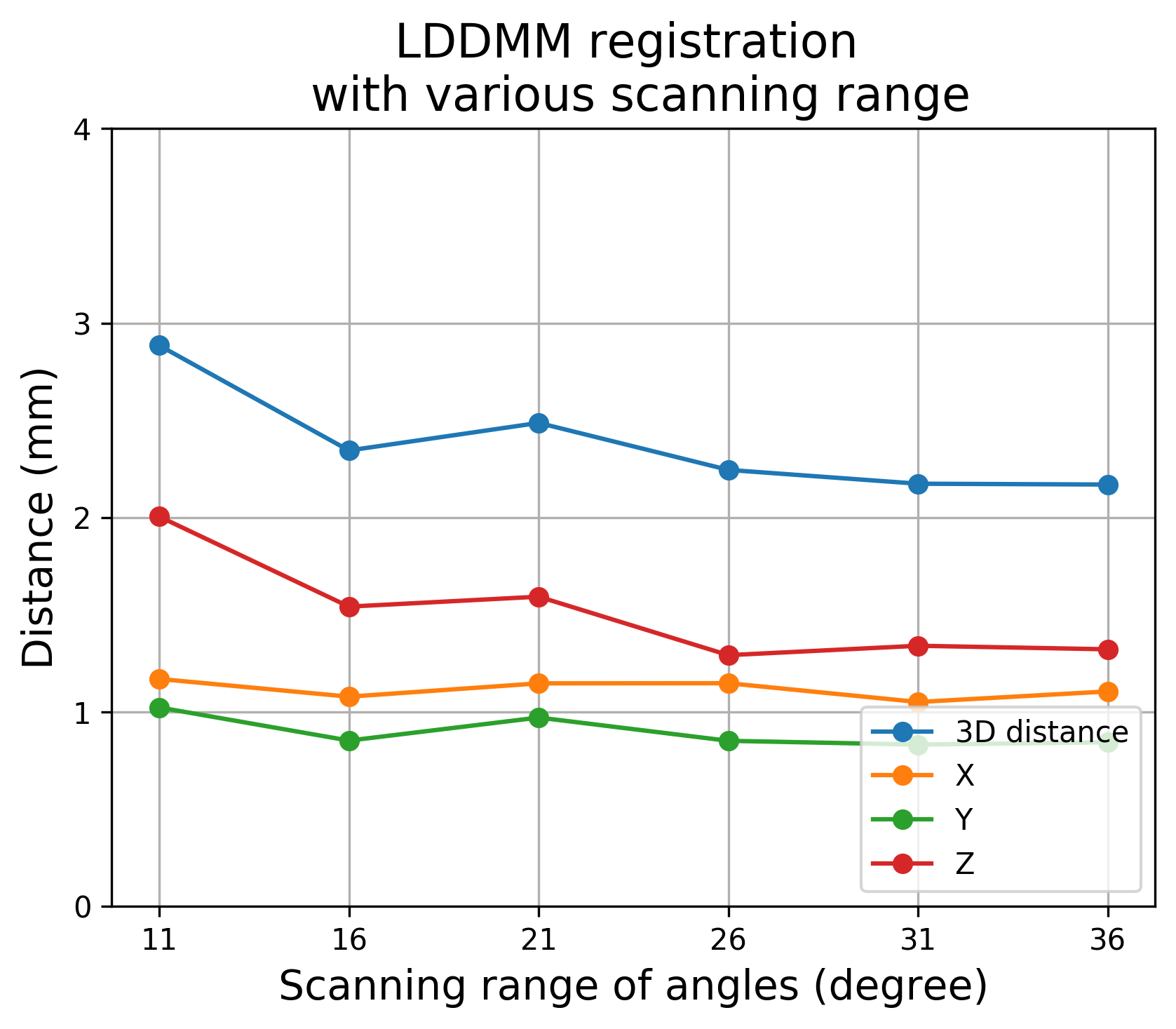} 
  \end{minipage} 
  \begin{minipage}[b]{0.248\textwidth}
    \centering
    \includegraphics[width=1\textwidth]{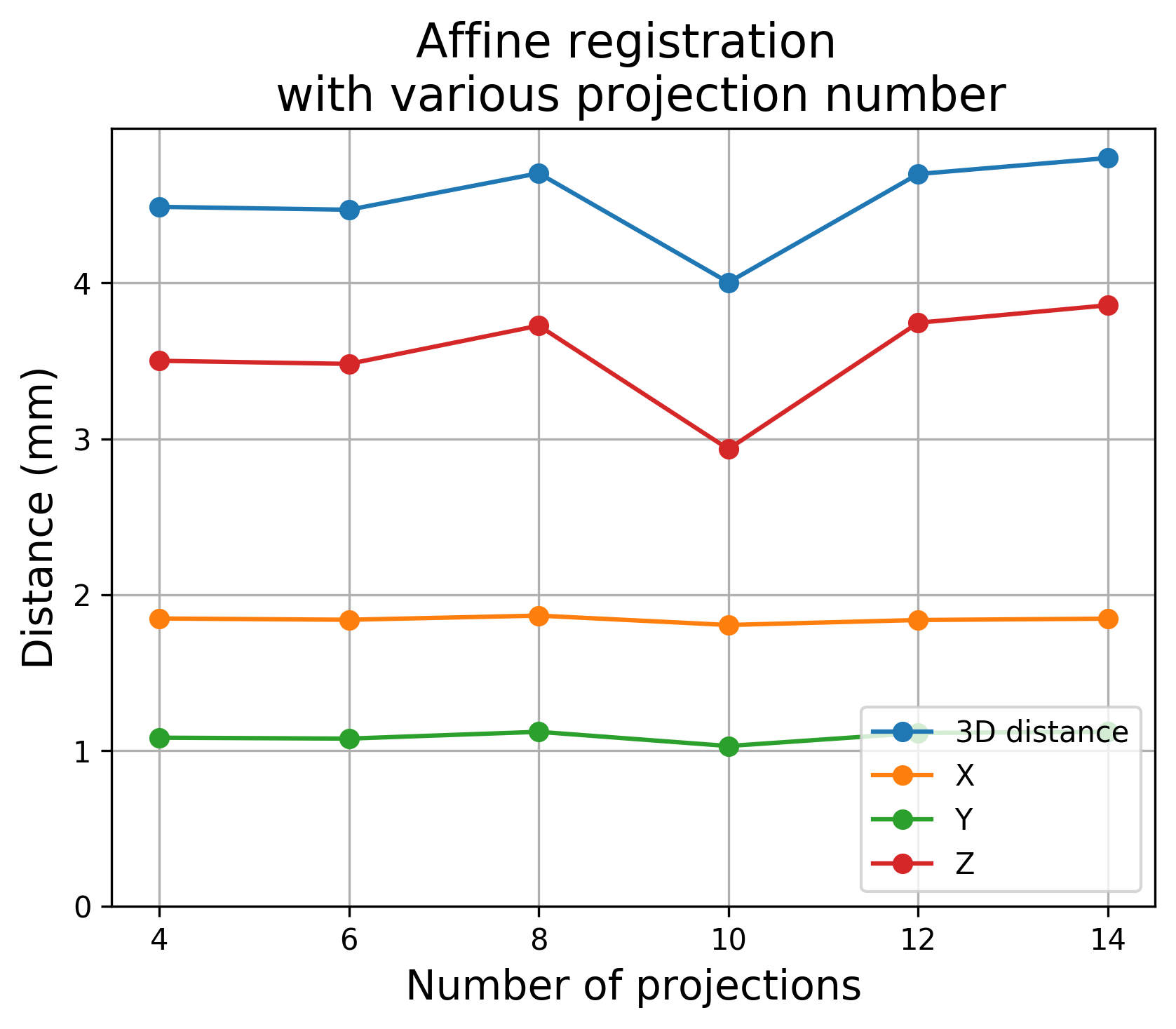} 
  \end{minipage}
  \begin{minipage}[b]{0.248\textwidth}
    \centering
    \includegraphics[width=1\textwidth]{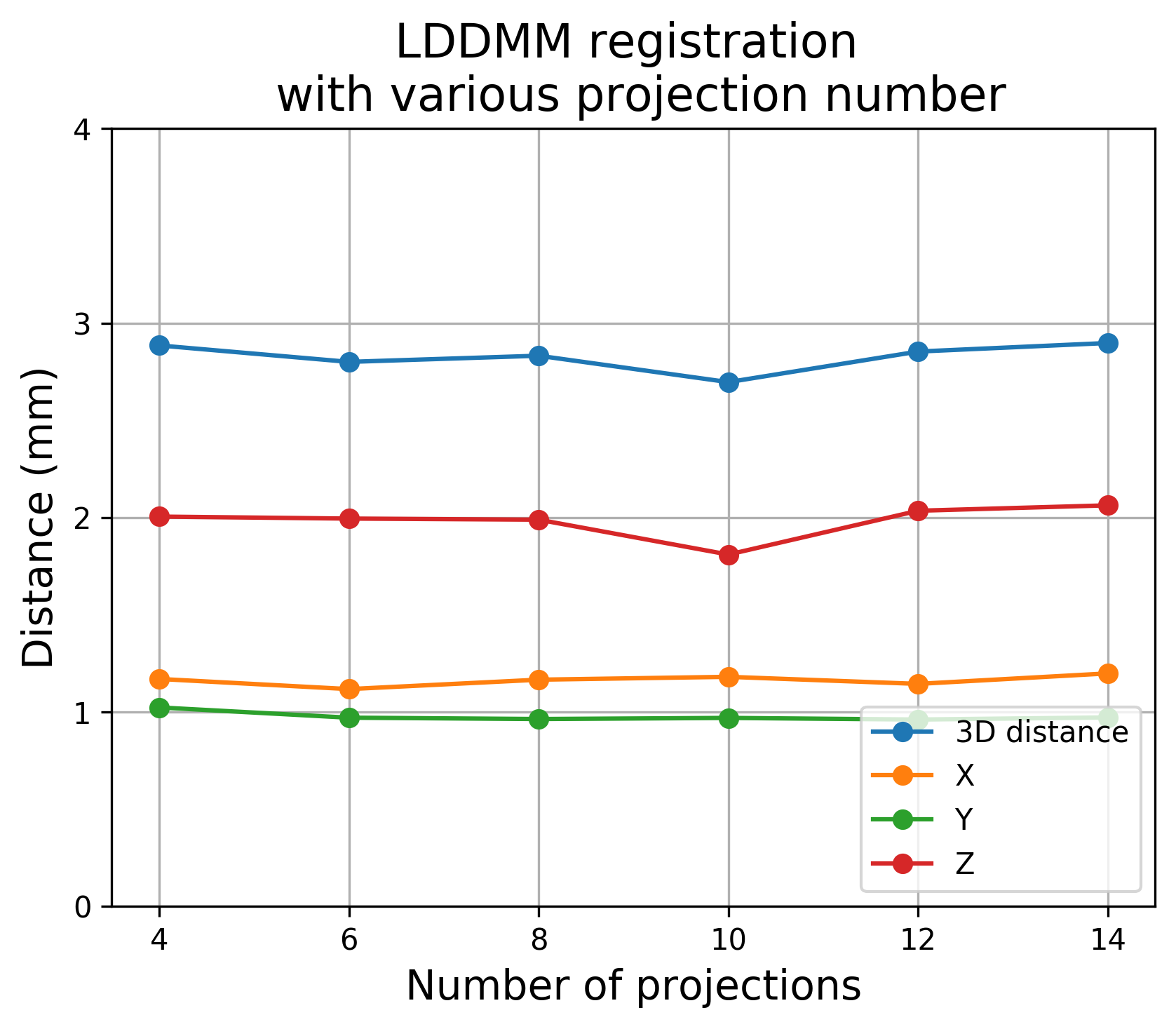} 
  \end{minipage} 
  \caption{Influence of angle range and number of projections. First 2 columns: Affine/LDDMM registrations accuracy improves with increasing angle range for 4 projections. Last 2 columns: Affine/LDDMM registration accuracies are relatively insensitive to the number of projections for an 11 degree angle range.}
  \label{fig_diff_angle_diff_proj} 
\end{figure}

Fig.~\ref{fig_diff_angle_diff_proj} shows that both affine and LDDMM registrations improve as the angle range increases. Note that improvements are most pronounced in the $Z$ direction, which is perpendicular to the receiver plane and thus has the least localization information in such a limited angle imaging setup. Fig.~\ref{fig_diff_angle_diff_proj} also shows that registration accuracy is relatively insensitive to the number of projections when restricting to a fixed angle range of 11 degrees. This indicates that angle range is more important than the number of projections. Further, when the radiation dose is a concern, larger scanning angles with fewer emitters are preferable. 

\section{Conclusion}
\label{sec:conclusions}

We proposed an sDCT-based DRR model and a 3D/2D framework for CT to sDCT registration. Our approach uses automatic differentiation and therefore easily extends to other tasks (e.g., 3D image reconstruction). We demonstrated that DRR-based 3D/2D LDDMM registration outperforms 3D/3D-reconstruction registration in a limited angle setting with small numbers of projections. Future work will extend our approach to an end-to-end 3D/2D deep network registration model. We will also explore alternative image similarity measures and quantitatively validate the approach on real CT/sDCT pairs.

\section*{Acknowledgements}
Research reported in this work was supported by the National Institutes of
Health (NIH) under award number NIH 1-R01-EB028283-01. The content is solely
the responsibility of the authors and does not necessarily represent the official
views of the NIH.

\bibliographystyle{splncs04}
\bibliography{ref}

\end{document}